# Mechanical Engineers' Training in Using Cloud and Mobile Services in Professional Activity

Maryna Rassovytska[1] and Andrii Striuk[2]

[1, 2] Institute of Information Technologies and Learning Tools of NAES of Ukraine,
04060 Kyiv, M. Berlyns'koho St., 9

[1]`rassovitskayamarina@mail.ru`, [2]`andrey.n.stryuk@gmail.com`

**Abstract.** The purpose of this article is to identify mobile and cloud services of mechanical engineers' professional activity and the principles of their use in higher technical education. There have been defined the criteria for evaluation of the tools for educational and professional activities. On the basis of this criteria, more than 30 various cloud services and mobile applications have been analyzed. The analysis has shown that the use of Autodesk cloud services and their integration with cloud services Google is appropriate for professional and practical training of specialists in applied mechanics, and it promotes an effective development of mechanical engineers' ICT competence. The learning tools integrated system model was proposed.

**Key terms:** InformationCommunicationTechnology, ICTTool, Competence, Educational Process, Model

## 1 Introduction

Professional activity of mechanical engineers require intensive use of ICT. A widespread use of mobile and cloud-oriented ICT is especially important in design documentation, calculations, and management of complex projects. Thus, training competitive professionals today is not possible without skills in using the cloud and mobile technologies in engineering.

The purpose of this article is to identify mobile and cloud services of mechanical engineers' professional activity and the principles of their use in higher technical education.

## 2 The Mechanical Engineers Needs in the Development of ICT Competence

In order to determine mechanical engineers' needs in the development of ICT competence it was organized an interrogation among of experienced specialists and university professors. The respondents were asked to identify the 10-point scale relevance of different skills to use ICT. The survey results are presented in Table 1.

Table 1. The mechanical engineers needs in the development of ICT competence

| № | ICT skills | Average rating |
|---|---|---|
| 1. | Using computer-aided design (CAD) | 8,71 |
| 2. | Computer simulation | 8,64 |
| 3. | Using Internet resources | 8,12 |
| 4. | Using cloud and mobile technologies | 7,85 |
| 5. | Business communication | 7,42 |
| 6. | Technical calculations | 7,39 |
| 7. | Professional collaboration | 7,39 |
| 8. | Preparation of technical documentation | 7,33 |
| 9. | Using databases | 6,28 |
| 10. | General algorithms and programming skills | 5,85 |
| 11. | Using visual programming | 5,15 |

Analysis of the Ukraine higher educational institutions curricula revealed that not all actual competence develop in a sufficient level. Mechanical engineers get a good fundamental training in the profession. But there is a need to teach them to work effectively in a team, using cloud and mobile services for computer modeling and design.

## 3 Cloud and Mobile Services in Mechanical Engineers' Professional Activity

In our research, we tried to identify most appropriate cloud and mobile services to use in the mechanical engineers' training. The main criteria for selection of mechanical engineers professional activity mobile and cloud services are:
– functionality;
– availability;
– easy access from different devices;
– ability to integrate with other software;
– support for collaboration.

We have analyzed about 30 modern cloud services and mobile applications. The most famous of them are A360, Fusion 360, GstarCAD, DWG FastView-CAD Viewer, CAD Pockets, 3D CAD Models Engineering, CAD Assistant, Onshape, GrabCAD, GnaCAD etc.

### 3.1 GstarCAD

GstarCAD [1] is a CAD software platform, using the Open Design Alliance DWG libraries to read and write the DWG file format made popular by the AutoCAD CAD package. GstarCAD is a capable alternative to other well-known CAD packages on the market, and provides OpenDWG file compatibility, as well as an interface which is very similar to that of AutoCAD. The system includes GstarCAD Mechanical, GstarCAD Architecture, DWG FastView for Mobile (Android), DWG FastView for Mobile (iOS), DWG FastView for Windows. There software comes in "standard", "academic" and "professional" versions.

GstarCAD MC (Mobile Client) is a mobile terminal platform developed by Suzhou Gstarsoft Co., Ltd, which is fully compatible with DWG drawings and helps users to view 3D drawings in multi-dimension perspectives and view, edit, annotate and share 2D CAD drawing files in the mobile terminal. [2]

Key features of GstarCAD MC:

– offline drawings;

– font management;

– LISP extension;

– view 3D drawing, 360 degree continuous rotate supported;

– precise design (when you select the «Line, Polyline..» commands, the coordinates of the point can be displayed);

– cloud files service (Dropbox connection service).

### 3.2 CAD Pockets

CAD Pockets (formerly ZWCAD Touch) is a powerful mobile CAD app tailor-made for designers, architects and engineers [3]. It supports various file formats ranging from DWG, DXF, to DWF. You can instantly open files without any lag, regardless of size. Except viewing the drawings, you can also revise, markup, and export files to PDF or DWF and share the file with your colleagues via email. CAD Pockets includes many popular third-party cloud storage services like Google Drive, Dropbox, and so on. You can sync up your files to the cloud and keep your local files updated in real time, saving you the trouble of transmitting files via cable. You can conveniently communicate with designers or clients even if you are on the go, avoiding the chore of carrying clumsy laptops or paper drawings. CAD Pockets also supports offline usage. Simply by saving files into your Android device and you are free to go. Moreover, you can view files that have been downloaded to your cloud drive without access to the Internet.

### 3.3 PARTcommunity 3D CAD

The PARTcommunity 3D CAD models app is a download service for 3D CAD data by CADENAS [4]. This app provides engineers and purchasers in the mechanical engineering, automotive and building industry direct access to thousands of parts from more than 400 certified catalogs of leading global manufacturers. The 3D CAD models are compliant with major world standards and are suitable for use in current CAD systems, such as: CATIA, Autodesk Inventor, SolidWorks, Creo Parametric, NX, AutoCAD, Solid Edge, etc.

### 3.4 OPEN CASCADE CAD Assistant for Android

OPEN CASCADE CAD Assistant for Android is an offline viewer and converter for 3D CAD files and meshes. Basic functionality is provided by CAD Data Exchange component of Open CASCADE Technology (OCCT) [5].

The supported file formats and data are:
– BREP: native OCCT format for shape geometry, topology, and assembly structure;
– IGES (5.1 and 5.3): shape geometry, colors, top-level object names, file information;
– STEP (AP203 and AP214): shape geometry, assembly structure, colors, names, validation properties, file information.

Additional functionality is viewing of mesh models with associated data, implemented using the mesh visualization component of OCCT. The mesh can be read from file in PLY, STL, or OBJ format. STL format is de-facto standard in 3D printing. PLY format has capability to store data associated with mesh nodes and elements. If you have in your application a model represented by mesh, it can be saved easily to PLY format and viewed on a tablet. Additional data (RGB color or scalar) can be added to each polygon or node. OBJ format is oriented at 3D animation and supported by most 3D graphics applications.

CAD Assistant allows you to view the mesh in wireframe, shaded, and shrink view. If the mesh contains associated color or scalar data, it can be viewed with elements colored according to the selected quantity. For scalar quantities interactive color scale is shown, providing controls to manipulate the range of displayed quantities.

### 3.5 Onshape

Onshape [6] is the full-cloud professional 3D CAD application built for agile product design teams working together using an Android Phone, Android Tablet, iPhone, iPad, or a web browser. Onshape was built from scratch for the way today's engineers, designers, and manufacturers really work.

Oneshape's intuitive sketch tools:
– create concept sketches;
– create lines, arcs, circles, rectangles, splines, and more;
– modify sketch entities using trim, offset, mirror, pattern;
– add geometric constraints and dimensions to control the behavior of your sketches;
– modify your sketches by changing dimension values to get the best design;
– real-time constraint solving enables you to create and test mechanism layouts;
– use edges from existing 3D geometry;
– sketches update when references to surrounding geometry update.

### 3.6 GrabCAD

GrabCAD [7] mobile app offers access to both the GrabCAD public library of CAD models as well as private projects stored on GrabCAD Workbench.

Using the app you can:

– view CAD models in full 3D, regardless of the format they were created in (3D viewer only supported in Android 4.0+);
 – view and respond to comments that Workbench collaborators have made;
 – create Workbench projects and upload files to them;
 – track file changes in your Workbench project;
 – search for models in the Community and mark them as Liked;
 – get notifications about updates and comments on your Workbench projects.

### 3.7 PadCAD

PadCAD CAD Drafting is an easy to use CAD application designed for small drafting projects such as home additions, small remodeling projects and site surveys [8]. With PadCAD, anyone can produce clean, clear drawings and export them to a professional CAD application like AutoCAD. PadCAD was designed with ease of use, speed and mobility in mind and is not a full blown CAD application. PadCAD has a shallow learning curve and is specifically designed for people with little or no previous experience with CAD software or drafting applications. PadCAD does not require an internet connection, except when you export a drawing. PadCAD can export drawings as PDF, DXF or image files (PNG format).

### 3.8 GnaCAD

GnaCAD is a full-fledged CAD system, not inferior to their desktop counterparts [9]. Key features of GnaCAD:

 – there is no size limit of opening files;
 – opening drawings from the memory card of the device;
 – ability to work with drawings located directly in Google Drive;
 – adding comments and replies (if the drawing was stored in Google Drive);
 – support for 3D views (including users views);
 – photo-realistic images of 3D models using ray tracing;
 – support layouts and viewports.

### 3.9 ARES Touch

ARES Touch is the application available on Android to create and modify DWG drawings [10]. DWG drawings can be shared and synchronized between ARES Touch and any other CAD software via Dropbox, Box, iCloud or Google Drive but also simply by email or USB cable.

Some of the key features of ARES Touch include:

– DWG support: We use .dwg as our main format. You can create and modify drawings in this popular format. No import nor export, you keep the highest compatibility.

– The natural extension of your favorite CAD software: We built an intuitive touch user interface. It has been redesigned for touch but is very intuitive for any CAD user.

– High precision: Discover in our video how the loupe, entity snap, tracking and coordinates input help you to draw with the same precision as on the desktop.

– Advanced annotation tools: A full set of dimensioning tools and unique tools designed for mobile such as Picture Note (insert picture from camera with comments), Voice Note, Pack & share…

– 150+ commands: The most complete app ever seen for mobile. Get all the power to create and modify your drawings on the go.

– Advanced programming interface: ARES Touch includes a C++, Lisp, Tx and DCL API. It makes it very easy to migrate applications developed for the desktop.

### 3.10 CADianAnyView

CADianAnyView is a Mobile CAD Program that supports DWG (up to AutoCAD 2013) as it opens DWG/DXF, and you can view and edit a drawing anytime and anywhere [11]. Quality improvement, delivery date reduction, and cost reduction are anticipated by the accumulation of these kinds of opinions since customers, employers, engineers, reviewers, and distribution suppliers can analyze products and cooperate with each other from product development to the end of product life.

CADianAnyView supports as follows:
– snap function, zoom, pan, 3D orbit function;
– line, polyline, circle text, free curve drawing;
– measurement function (area, distance, size);
– editing function(erase, move, copy);
– layer function;
– view sight at various angles.

### 3.11 SchemataCAD

In SchemataCAD [12] viewer you can easily view 2D CAD drawings stored on your tablet or mobile phone. It is also possible to open a drawing directly from an email attachment, web page or from "file manager".

Viewer opens CAD file formats:
– DWG (up to the latest version 2017 - AC1027);
– DXF (all versions);
– SCH (format of software SchemataCAD).

Viewer contains AutoCAD SHX standard fonts. Also are accepted other SHX fonts and shapes, which are in same folder as is a DWG/DXF file (or define folder with SHX fonts in configuration).

### 3.12 SketchUp Viewer

The SketchUp Viewer [13] brings 3D models to life on your Android phones and tablets, allowing you to open and view SketchUp models any time, anywhere, on the device you want to view them on. Explore and share 3D models:

– download models to your device from your 3D Warehouse, Trimble Connect or Dropbox account for seamless offline viewing;

– open SketchUp models directly from email attachments, or open files from other cloud service apps like Google Drive;

– navigation features include multi-touch gestures for Orbit, Look Around, Pan, Zoom, and Zoom Extents, as well as a Camera menu with options for toggling between Perspective and Orthographic camera modes, and adjusting the Field of View;

– use the Scenes menu to select from any of the standard camera views (top, side, bottom, etc.) or access the custom Scenes that you created in SketchUp. The app supports the following scene properties: Camera location and properties, Hidden Geometry, Visible Layers, Active Section Planes, Standard Edge Styles, Face Styles, Background/Sky/Ground Style settings, Watermarks and Axes Location.

– use the Layers panel to toggle the contents of your SketchUp model layers on or off;

– use the View panel to choose from any of SketchUp's Face styles and to toggle X-Ray mode on or off. The view menu also offers visibility toggles for Hidden Geometry, Section Planes, Section Cuts and Axes;

– use the Tape Measure tool to quickly measure objects in your SketchUp model.

### 3.13 Autodesk A360

Autodesk A360 [14] was designed specifically for architects, engineers and designers to comment, markup and collaborate 2D & 3D CAD models. With over 100 CAD file and additional file formats supported, A360 allows you to upload and view any file you have, no matter what software you used to create it. Whether you are at the office, or doing fieldwork, you can take A360 with you and stay up-to-date with your projects.

View 2D & 3D CAD models:

– view over 50 different CAD file formats including: AutoCAD (DWG), DWF, Inventor (IPT, IAM, IDW), Revit (RVT), SolidWorks (SLDPRT, SLDASM, ASM), Navisworks (NWD, NWC), CATIA (CATPART, CATPRODUCT), Fusion 360 (F3D) and more;

– upload and view CAD models from email attachments;

– upload and view CAD models from device's local storage, Dropbox, Box, Buzzsaw, OneDrive etc.

Navigate large-scale 2D & 3D CAD models:

– isolate and present object properties;

– navigate model parts and layers;

– measure the distance, angle or area between points in your drawing;

– intuitive touch-based navigation including: zoom, pan, orbit and rotate 3D models;

Collaborate with your clients, colleagues and others all-in-one place:

– review and markup your designs for easy collaboration;
– comment directly on your designs and keep track of changes;
– take and upload photos directly from the device to your account to document work progress;
– invite new members to join your project in progress and collaborate on designs;
– share any CAD file type directly from your Android device, including: AutoCAD (DWG), DWF, Solidworks, Revit, CATIA and more.

### 3.14 AutoCAD 360

AutoCAD 360 [15] is a free DWG viewing application, with easy-to-use drawing and drafting tools that allow you to view, create, edit, and share AutoCAD drawings across web and mobile devices - anytime, anywhere. Simplify your site visits with the most powerful drafting and editing tool available.

The AutoCAD 360 mobile app offers an abundance of features and capabilities. Upload and open 2D DWG drawings directly from email or external storage and view all aspects of your DWG file, including external references, layers, and image underlays. Upgrade to AutoCAD 360 Pro to enable editing and drawings tools. Whether working online or offline, in the office or in the field, design every detail, everywhere.

Whether you want to draft, view or mark up a DWG file, AutoCAD 360 has the plan that's right for you.

Features of AutoCAD 360:
– open and view your DWG files;
– measure accurately while on site;
– view a drawing's coordinates;
– use multi-touch zoom and pan to easily navigate large drawing;
– work offline and sync your changes once back online;
– external cloud storage connectivity (Google Drive, Dropbox, OneDrive, and more);
– use GPS to orient yourself within a drawing;
– improve communication by adding comments and images and invite responses using the Design Feed;
– share your designs with others directly from mobile;
– use the free companion web app to easily access drawings from web browsers.

AutoCAD 360 Pro subscription advantages:
– new drawing creation;
– support for larger files and increased storage capacity;
– zll drawing and editing tools, including advanced tools, such as arc, offset and more;
– draw and edit shapes with accuracy using object snap and new keypad feature (keypad available on iPad only);
– select, move, rotate, and scale objects;
– editing capabilities are also available directly from external cloud storage files;
– add and edit text annotations directly on your drawing; no need for paper mark-ups;

- advanced annotation tools, including: cloud, mark up, arrow and more;
- advanced layer management;
- view and edit object properties;
- block palette containing all the existing blocks from the drawing, to allow the user to insert blocks;
- additional drawing tools and ability to view drawing coordinates;
- personal customer support via priority email channel.

### 3.15 Fusion 360

Fusion 360 is a cloud-based 3D CAD, CAM, and CAE platform for product development [16]. It combines industrial and mechanical design, simulation, collaboration, and machining in a single package. The tools in Fusion 360 enable fast and easy exploration of design ideas with an integrated concept-to-production toolset.

Fusion 360 is natively written for both Mac and PC, allowing you to use your preferred OS or both at no additional cost.

Autodesk Fusion 360 for Android lets you collaborate on 3D designs with anyone inside or outside your company. With the Fusion 360 app, you have the flexibility to view, mark up, comment, and collaborate on your Fusion 360 CAD models – anytime, anywhere. The app supports more than 100 file formats including DWG, SLDPRT, IPT, IAM, CATPART,IGES, STEP, STL, making it easy to share designs with your team, clients, partners, and friends. The free app works in conjunction with its companion cloud-based desktop product, Autodesk Fusion 360, a 3D CAD, CAM, and CAE tool for product design and development.

Features of Autodesk Fusion 360 for Android:
- upload and view more than 100 data formats including SLDPRT, SAT, IGES, STEP, STL, OBJ, DWG, F3D, SMT, and DFX;
- view and track project activities and updates;
- review large and small 3D designs and assemblies;
- measure the distance, area or angle between points in your 2D drawings or your 3D designs;
- access design properties and complete parts lists;
- isolate and hide components in the model for easy viewing;
- navigate by touch with zoom, pan, and rotate;
- review & mark up your designs for easy collaboration;
- comment directly on your designs and keep track of changes;
- upload photos to share info or report project status;
- comment on project activities;
- share with stakeholders inside and outside your company;
- share screenshots of the design with markups directly from the app.

## 4 The Results of Expert Assessment

The results of expert assessment of these tools are presented in Table 2. In addition, we took into account user reviews of these tools (according to the Google Play Market).

**Table 2.** Expert review of cloud and mobile services of mechanical engineers' professional activity

| Applications and services | Functionality | Availability | Easy access | Ability to integrate | Support for collaboration | Overall Rating |
|---|---|---|---|---|---|---|
| Fusion 360 | 9,5 | 8,8 | 8,9 | 10 | 10 | 47,2 |
| A360 | 9,8 | 8,7 | 8,2 | 10 | 10 | 46,7 |
| AutoCAD 360 | 8,8 | 8,7 | 8,8 | 9,7 | 9,8 | 45,8 |
| GstarCAD | 8,3 | 8,2 | 8,8 | 9,1 | 8,5 | 42,9 |
| CAD Pockets | 8,2 | 7,4 | 7,5 | 8,5 | 9,2 | 40,8 |
| GnaCAD | 8,6 | 8,5 | 8,2 | 7,1 | 8,4 | 40,8 |
| ARES Touch | 9,8 | 7,8 | 7,5 | 6,5 | 9,1 | 40,7 |
| Onshape | 9,1 | 8,3 | 8,4 | 5,1 | 9,2 | 40,1 |
| CADianAnyView | 7,9 | 8,1 | 6,4 | 6,5 | 8,0 | 36,9 |
| SketchUp Viewer | 4,8 | 8,6 | 9,1 | 5,1 | 7,4 | 35,0 |
| SchemataCAD | 3,5 | 9,8 | 9,9 | 6,4 | 5,2 | 34,8 |
| GrabCAD | 5,2 | 7,3 | 8,9 | 5,5 | 6,8 | 33,7 |
| PARTcommunity 3D CAD | 5,3 | 8,1 | 8,0 | 7,2 | 5,1 | 33,7 |
| PadCAD | 6,1 | 8,5 | 8,4 | 6,0 | 4,1 | 33,1 |
| CAD Assistant | 4,8 | 8,5 | 9,0 | 3,2 | 4,6 | 30,1 |

## 5 Integrated Use of Autodesk Products in the Training

The considered software products offer enough features to work with drawings. But they all are oriented to support formats and technologies implemented by Autodesk, so we pay more attention to software products offered by this company. Their wide range of tools enables comprehensive use of desktop software, cloud services and mobile applications with automated design and collaborate on design-project (Fig. 1).

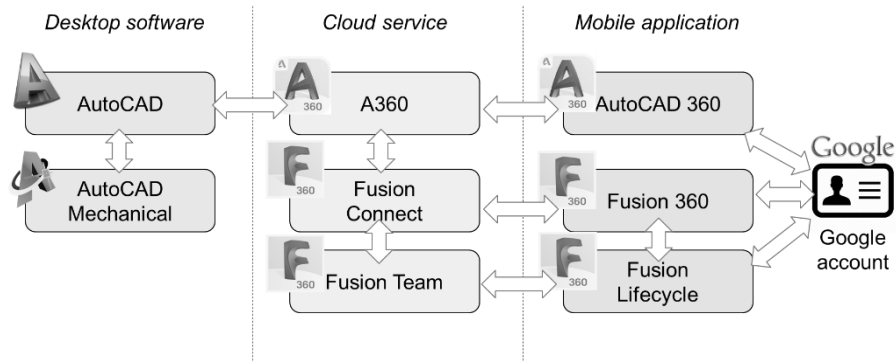

**Fig. 1.** Integrated use of Autodesk products in the training

## 6    Learning Tools Integrated System Model

Among all the possibilities of this complex application it showed be mentioned support for authentication via Google Account and integration of cloud services provided by this company. Thus, students have an opportunity to update knowledge and skills obtained during their study of fundamental disciplines and to use learning tools as an integrated system (Fig. 2).

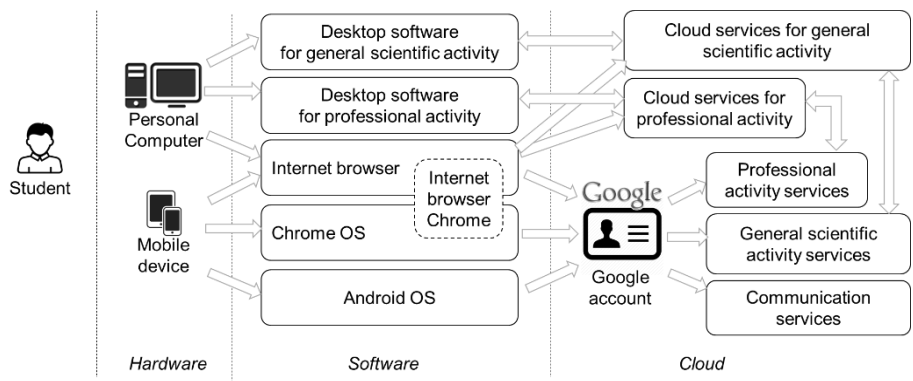

**Fig. 2.** Learning tools integrated system model

This model involves the use of various hardware and software available to users at a time. For the access to all necessary services it is enough to apply any Internet browser, but both special desktop and mobile applications can be used.

Cloud services for general scientific and professional activities can be used both independently and together with Google cloud services. In the latter case, a Google Account is a single point of access to various services, mobile applications, cloud storage, and communication tools that provide both effective educational process organization and professional activity. Google services are widely used in the training of specialists in applied mechanics while teaching general disciplines, and the use of Autodesk software will be appropriate in terms of the development of the ICT competence of mechanical engineers.

## 7  Conclusions

Thus, we identified the need for the development of ICT competence for bachelor of applied mechanics, including skills and ability to adopt and use mobile and cloud-oriented tools in their professional activity. There have been defined the criteria for evaluation of the tools for educational and professional activities. On the basis of this criteria, more than 30 various cloud services and mobile applications have been analyzed. Some of them have been examined in detail. The analysis has shown that the use of Autodesk cloud services and their integration with cloud services Google is appropriate for professional and practical training of specialists in applied mechanics, and it promotes an effective development of mechanical engineers ICT competence. The proposed learning tools integrated system model can be applied to vocational and practical training.